\documentclass[%
 aip,
 twocolumn,
 amsmath,amssymb,
 reprint,%
]{revtex4-1}

\usepackage{graphicx}
\usepackage{dcolumn}
\usepackage{hyperref}
\usepackage{siunitx} 
  \DeclareSIUnit\sq{\ensuremath{\Box}}
  \DeclareSIUnit\torr{Torr}
\DeclareUnicodeCharacter{2009}{\,}
\usepackage[T1]{fontenc}
\usepackage{mathptmx}
\usepackage{etoolbox}
\usepackage{caption}

\makeatletter
\def\@email#1#2{%
 \endgroup
 \patchcmd{\titleblock@produce}
  {\frontmatter@RRAPformat}
  {\frontmatter@RRAPformat{\produce@RRAP{*#1\href{mailto:#2}{#2}}}\frontmatter@RRAPformat}
  {}{}
}%
\makeatother
\begin{document}

\preprint{AIP/123-QED}

\title[]{Large active-area superconducting microwire detector array with single-photon sensitivity in the near-infrared}
\author{Jamie S. Luskin}
\thanks{These two authors contributed equally: J.S.L., E.S.}
\affiliation{Jet Propulsion Laboratory, California Institute of Technology, Pasadena, CA 91109, USA}
\affiliation{University of Maryland, College Park, MD 20742, USA}

\author{Ekkehart Schmidt }

\altaffiliation{Now at: Robert Bosch GmbH, Reutlingen, Germany}

\author{Boris Korzh}
\email{Corresponding Author: jamie.s.luskin@jpl.nasa.gov}

\author{Andrew D. Beyer}
\author{Bruce Bumble}
\author{Jason P. Allmaras}
\author{Alexander B. Walter}
\author{Emma E. Wollman}
\affiliation{Jet Propulsion Laboratory, California Institute of Technology, Pasadena, CA 91109, USA}

\author{Lautaro Narv\'{a}ez}
\affiliation{Division of Physics, Mathematics and Astronomy, California Institute of Technology, Pasadena, CA 91125, USA}

\author{Varun B. Verma}
\author{Sae Woo Nam}
\affiliation{National Institute of Standards and Technology, Boulder, CO 80305, USA}

\author{Ilya Charaev}
\affiliation{Research Laboratory of Electronics, Massachusetts Institute of Technology, Cambridge, MA 02139, USA}
\affiliation{University of Zurich, Zurich 8057, Switzerland}
\author{Marco Colangelo}
\author{Karl K. Berggren}
\affiliation{Research Laboratory of Electronics, Massachusetts Institute of Technology, Cambridge, MA 02139, USA}

\author{Cristi\'{a}n Pe\~{n}a}
\affiliation{Fermi National Accelerator Laboratory, Batavia, IL 60510, USA}

\author{Maria Spiropulu}
\affiliation{Division of Physics, Mathematics and Astronomy, California Institute of Technology, Pasadena, CA 91125, USA}

\author{Maurice Garcia-Sciveres} 
\author{Stephen Derenzo}
\affiliation{Lawrence Berkeley National Laboratory, Berkeley, CA 94720, USA}

\author{Matthew D. Shaw}
\affiliation{Jet Propulsion Laboratory, California Institute of Technology, Pasadena, CA 91109, USA}

\date{\today}

\begin{abstract}
Superconducting nanowire single photon detectors (SNSPDs) are the highest-performing technology for time-resolved single-photon counting from the UV to the near-infrared. The recent discovery of single-photon sensitivity in micrometer-scale superconducting wires is a promising pathway to explore for large active area devices with application to dark matter searches and fundamental physics experiments. We present 8-pixel 1mm\textsuperscript{2} superconducting microwire single photon detectors (SMSPDs) with \SI{1}{\micro\metre}-wide wires fabricated from WSi and MoSi films of various stoichiometries using electron-beam and optical lithography. Devices made from all materials and fabrication techniques show saturated internal detection efficiency at 1064 nm in at least one pixel, and the best performing device made from silicon-rich WSi shows single-photon sensitivity in all 8 pixels and saturated internal detection efficiency in 6/8 pixels. This detector is the largest reported active-area SMSPD or SNSPD with near-IR sensitivity published to date, and the first report of an SMSPD array. By further optimizing the photolithography techniques presented in this work, a viable pathway exists to realize larger devices with cm\textsuperscript{2}- scale active area and beyond.  
\end{abstract}

\maketitle

\section{\label{sec:level1}Introduction}
 Superconducting nanowire single photon detectors (SNSPDs) are a leading detector technology for time-correlated single photon counting from the UV to the near-infrared. They have become essential components in optical communication~\cite{DSOC, LLCD, Li:20}, Quantum Information Science ~\cite{QI1, QI2, QI3, QI4, QI5, QI6, QI7, QI8}, and LIDAR~\cite{imaging1, Lidar2, Lidar3, Lidar4} experiments due to their unique combination of high timing resolution and high efficiency.
 
 In recent years, the active areas of SNSPDs have steadily increased from the scale of a single-mode optical fiber diameter ($\sim$\SI{100}{\micro\meter\squared}) towards the \SI{}{\milli\meter\squared} regime~\cite{kilopixel_array, Wollman2021, LargeSNSPD}.  As new applications for SNSPDs that leverage these large active areas emerge, there is a growing interest in exploring novel materials and device architectures. For applications in high energy physics such as dark matter (DM) searches, it is necessary to realize SNSPDs with larger active areas than the current state-of-the-art, by at least an order of magnitude. Scaling devices beyond the \SI{}{\milli\meter\squared} regime remains an open question that presents many challenges in terms of design, fabrication, and readout. 
 
 Traditional SNSPDs are comprised of nanowires with widths $\mathcal{O}$(\SI{100}{\nano\meter}) meandered over the photoactive area and are fabricated using electron-beam lithography (EBL) to achieve the required resolution and optimal performance. Recent advances in the preparation of thin superconducting films have enabled the fabrication of a novel class of single-photon detectors with wire widths of several micrometers. We refer to these as superconducting microwire single-photon detectors (SMSPDs). In the near-IR, single-photon sensitivity in SMSPDs was first shown in niobium nitride (NbN) microstrips~\cite{Korneeva2018} and shortly after in MoSi microstrip structures~\cite{micron1}. Following this, large area meander devices with saturated internal detection efficiency at \SI{1550}{\nano\meter} were demonstrated in both silicon-rich tungsten silicide (WSi)~\cite{ChilesSMSPD2020} and molybdenum silicide (MoSi)~\cite{Charaev2020}, where the wire widths were on the order of 1-3~\SI{}{\micro\meter}. UV detection at an elevated temperature of 3.4~K has also been demonstrated for large area MoSi SMSPDs~\cite{Lita2021} and TaN SMSPDs have been developed for X-ray detection~\cite{XSMSPD}. The timing resolution in NbN strips with widths up to \SI{3}{\micro\meter}~\cite{NbNMicrostrip} has been studied in a broad range of wavelengths and remains an important topic of fundamental research due to the complex role of vortex dynamics in the detection mechanism. Additionally, micron-scale wires are amenable to patterning with broadly available photolithography, which enables stable wafer-scale fabrication of devices. Direct laser writing has already been explored for micron-wide wires~\cite{Protte2022}, while the use of a deep-UV stepper is introduced in this work. Because of these developments, SMSPDs are promising to explore for \SI{}{\centi\meter\squared}-scale arrays in the future.


\begin{figure*}
    \centering
    \includegraphics[width = 1\linewidth]{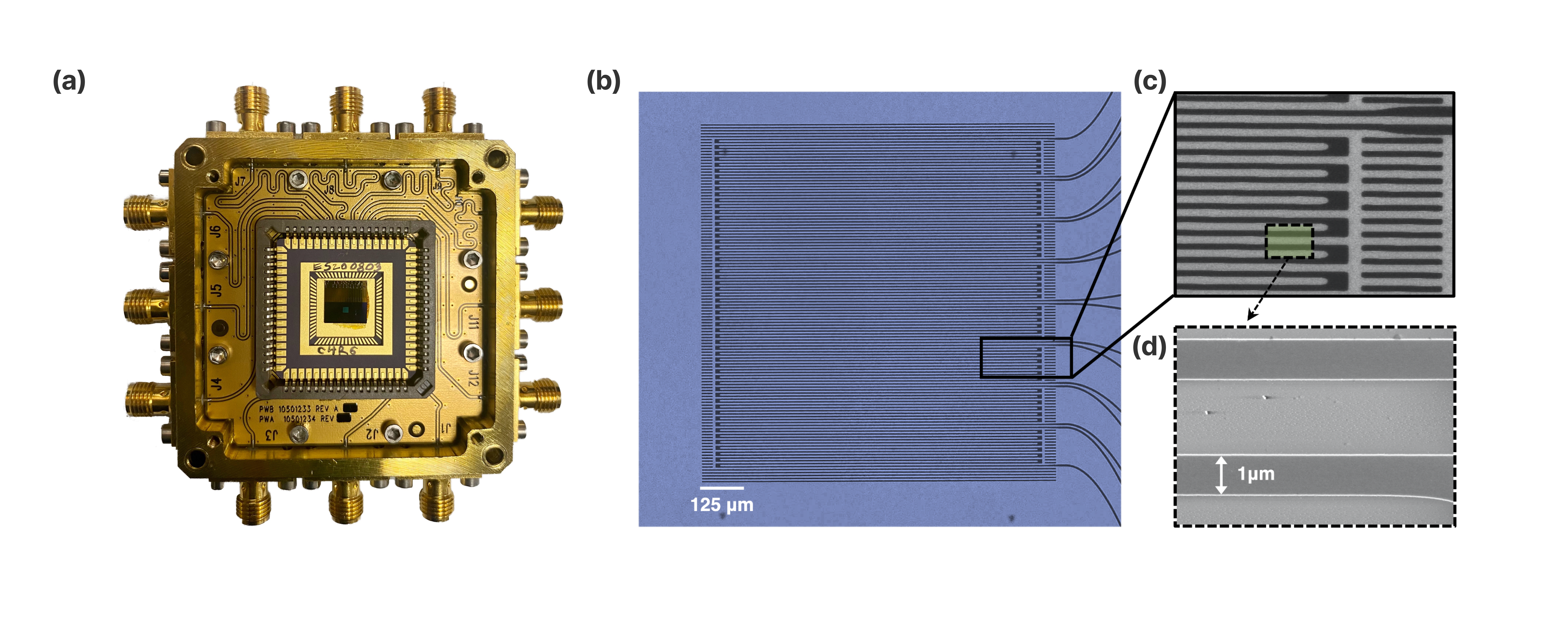}
    \caption{(a) Photograph of the array in detector packaging (b) Optical image of the detector chip (c) scanning electron micrograph of the meander (d) SEM showing zoomed in view of a micron-wide wire made with photolitography}
    \label{fig:my_label2}
\end{figure*}
 
In this work, we investigate the feasibility a variety of materials for large active-area SMSPDs, including WSi of various stoichiometries and MoSi. An 8-pixel SMSPD array with meander geometry, \SI{1}{\micro\meter}-wide wires, 25$\%$ fill factor, and a total active area of 1 mm\textsuperscript{2} was fabricated from each material for comparison. The devices fabricated from WSi and MoSi films were patterned using EBL and photolithography, respectively. This work paves the way to realize cm\textsuperscript{2}-active area devices which can have transformative capabilities in high energy physics. 

The \SI{}{\milli\meter\squared} arrays fabricated in this work are a promising photosensor for a novel DM search experiment using cryogenic scintillating targets. Scintillating \textit{n}-type GaAs doped with silicon and boron has recently been proposed as a target for detecting sub-GeV DM via electron recoils~\cite{PhysRevD.96.016026,Derenzo2018,DERENZO2021164957}. The emission bands of \textit{n}-type GaAs are centered at 860 nm, 930 nm, 1070 nm, 1335 nm, with the 930 nm
band having the highest luminosity. This application requires coupling the target to a single-photon detector with both near-IR sensitivity and a large active area to achieve high optical coupling efficiency. With this application in mind, the devices in this work were characterized at $\lambda = 1064$~nm. The photoresponse at \SI{1064}{nm} and dark count rates (DCR) were characterized at 0.8~K. Saturated internal detection efficiency at $\lambda = 1064$~nm  was achieved for all devices. The same mm$^{2}$ array makes an ideal photosensor for new searches for axion- and dark photon-DM such as the BREAD experiment~\cite{BREAD}.
The proposed BREAD design consists of a cylindrical metal barrel for broadband conversion of DM into photons with an internal parabolic reflector that focuses the converted photons onto a photosensor with an active area of approximately 1~mm$^{2}$. The current device with saturated internal detection efficiency at $\lambda = 1064$~nm is able probe the unexplored region of parameter space with DM masses around 1.2~eV while meeting the active area requirements.

\section{Materials}
Recent work by Xu et. al.~\cite{xu2021superconducting} demonstrates that NbN can be used in combination with an optical cavity to achieve a high system detection efficiency in small active area SMSPDs. WSi and MoSi are also promising materials for large active-area detectors since the films are amorphous and are thus expected to be very homogeneous. 
 
 A long saturation plateau in the photon count rate (PCR) curve allows for device operation at a bias current ($I_{B}$), much lower than the detector critical current ($I_{C}$), which decreases the amount of intrinsic detector dark counts. This can be achieved in a device by decreasing the density of charge carriers in the material, or by decreasing the superconducting energy gap of the material~\cite{Colangelo2022}. Decreasing the density of states reduces the number of Cooper pairs that need to be broken in order to weaken the superconducting state sufficiently for the nucleation of vortices. Once nucleated, vortices traverse the microwire and dissipate energy, which ultimately breaks the superconductivity across the whole microwire and leads to a detection event~\cite{Vodolazov2017}. This enhancement in sensitivity can be achieved by increasing the material resistivity or by reducing the thickness of the film. Decreasing the superconducting gap of a material increases the number of quasiparticles that can be excited by a photon of a given energy. The gap energy is directly proportional to the critical temperature ($T_{C}$) of the device. In WSi, increasing the silicon content of the film both increases the resistivity and reduces $T_{C}$~\cite{ChilesSMSPD2020,verma2020singlephoton, Colangelo2022}.
 
 One drawback of high resistivity materials is that the increased resistivity results in a larger kinetic inductance ($L_K$) of the material, which makes the recovery timescale of the device slower ($\tau_R = L_K/R_L$, where $\textit{R}_L$ is the load resistance). Additional challenges introduced by lowering $I_{C}$ of the device include a decreased output signal level, and a reduced operation temperature, which increases the level of complexity of the cryogenics. 
 
 We investigated 8-pixel meander SMSPD arrays made from three WSi films with different stoichiometries and thicknesses, and one made from a MoSi film. The sheet resistances were measured using a 4 probe setup at room temperature. The $T_{C}$ was defined as the temperature at the inflection point of the measured resistance–temperature curve. The composition of Film 1 was determined using Secondary Ion Mass Spectroscopy (SIMS), while Film 2 and 3 were analyzed using Rutherford Backscattering Spectrometry (EBS). The properties of the films are listed in table 1.  

\begin{table*}
\caption{\label{tab:table3}Properties of the films used for detector nanofabrication}
\begin{ruledtabular}
\begin{tabular}{ccccccccc}
 Name  & Substrate & Film & Stoichiometry (W:Si) & Thickness & Sheet Resistance & $\rho_{300}$ & $T_C$ & Fabrication  \\ 
\hline

Film1 & Si & WSi & 52:48 \SI{\pm10}{\percent}  & \SI{2.2}{\nano\meter} & \SI{1000}{\ohm} & \SI{220}{\micro\ohm\centi\meter} & \SI{2.6}{\kelvin} &  e-beam lithography\\  
Film2 & $\mathrm{SiO_2}$         & WSi          & 70:30 \SI{\pm1}{\percent} & \SI{5.1}{\nano\meter} & \SI{570}{\ohm} & \SI{177}{\micro\ohm\centi\meter} & \SI{3.25}{\kelvin} & e-beam lithography \\
Film3 & $\mathrm{SiO_2}$  & WSi    & 42:58 \SI{\pm1}{\percent}& \SI{4.7}{\nano\meter} &          \SI{750}{\ohm} & \SI{320}{\micro\ohm\centi\meter} & \SI{1.9}{\kelvin} & e-beam lithography \\
Film4 & $\mathrm{SiO_2}$         & MoSi            & unknown  & \SI{2}{\nano\meter} &             \SI{1180}{\ohm} & \SI{236}{\micro\ohm\centi\meter}  & \SI{3.4}{\kelvin} & photolithograpy\\       
\end{tabular}
\end{ruledtabular}
\end{table*}
Film 1 was co-sputtered from tungsten and silicon targets using the procedures outlined in Ref.~\cite{ChilesSMSPD2020}. 
Films 2, 3 and 4 were deposited onto an oxidized Si(100) substrate with a \SI{240}{\nano\meter}-thick thermal oxide. 
Film 2 was sputtered from a $\textrm{W}_{50}\textrm{Si}_{50}$ target in an Ar atmosphere at a pressure of \SI{5}{\milli\torr} and a sputter power of \SI{130}{\watt}. Film 3 was sputtered from a $\textrm{W}_{30}\textrm{Si}_{70}$ target in an Ar atmosphere at a pressure of \SI{5}{\milli\torr} and a sputter power of \SI{130}{\watt}. 
Film 4  was made by co-sputtering process from Mo and Si targets in an argon atmosphere at a pressure of 2.5~mTorr. The Mo target was DC biased with a current of 130~mA, while 73~W of RF power was applied to the Si target. 

\section{Design and Fabrication}
The SMSPDs in this work were designed as \SI{1}{\micro\meter}-wide meanders with a \SI{1}{\milli\meter\squared} active area and a fill factor of 0.25. The area was chosen to be as large as possible while still fitting into a single write field of the EBL tool, to avoid stitching errors. The device was not put into a cavity, but a natural future extension of this work would be to embed the microwires in a standard, front-side illuminated cavity in order to increase the detection efficiency.

To investigate uniformity and yield, the detector was divided into \num{8} pixels, each with an active area of $1 \times 0.125$\SI{}{\milli\meter}. The pixels were designed with contact pads on either side, which enables the interconnection of all pixels in an arbitrary fashion and also allows for the use of differential readout~\cite{Colangelo2021} if desired. In this work, we connected one side of each pixel to GND and the other to an individual readout line. In this single-ended configuration, the timing jitter of the Film 3 device was measured to be \SI{325}{\pico\second}. Details on this measurement can be found in the supplementary information.

We patterned microwires with using EBL (Films 1,2,3) and optical lithography (Film 4). The EBL was done with the negative tone ma-N2401 resist at an acceleration voltage of \SI{48}{\kilo\electronvolt} and a dose of \SI{330}{\micro\coulomb\per\centi\meter\squared}. The development was performed using AZ 300 MIF developer. The optical lithography was performed using a deep-UV stepper with a wavelength of $\lambda = \SI{248}{\nano\meter}$ in off-axis alignment mode. In addition, a backside anti-reflection coating (BARC) was used to prevent unwanted reflections from the wafer, which can cause a widening or constriction of the patterned wires. More details on the optical lithography of SNSPDs can be found in \cite{BeyerOffAxis2015}. In addition, dummy wires were added to either side of the active area for proximity effect correction to allow for a homogeneous exposure of the microwire pattern. The pattern was transferred via inductively-coupled plasma reactive-ion etching using $\textrm{O}_2$ and $\textrm{CF}_4$. After etching, the microwires were covered with \SI{50}{\nano\meter} of $\textrm{SiO}_2$ for passivation.

\section{Readout}
Typical readout of SNSPDs is based on room temperature, low-noise RF amplifiers with AC-coupling~\cite{Marsili2013}. An AC-coupled readout can prevent SNSPDs from reaching their maximum $I_C$ due to an early onset of latching~\cite{Kerman2013}. If the recovery timescale of the SNSPD is significantly longer than the timescale corresponding to the low-frequency cut-off of the amplifier, there is a significant undershoot in the readout signal, which results in additional current being sent to the detector during the recovery phase, causing the detector to latch. The low frequency cut-off of low-noise RF amplifiers is typically $\mathcal{O}$(\SI{100}{\kilo\hertz}), which is compatible with SNSPD recovery times $\mathcal{O}$(\SI{100}{\nano\second}). In the case of the \SI{1}{\micro\meter}-wide wire devices in this work, the sheet-inductance is 650~pH/square and $L_K = 17.5~\mu$H for the device, thus the reset timescale is $\mathcal{O}$(\SI{1}{\micro\second}). To overcome this limitation, we developed a DC-coupled amplifier based on a HEMT first stage, similar to the one described in Ref.~\cite{Kerman2013}, operating at \SI{40}{\kelvin}. With the addition of a second stage SiGe amplifier, the total gain at \SI{40}{\kelvin} is \SI{40}{\decibel} with a nominal \SI{3}{\decibel} bandwidth of 600~MHz. An overview diagram of the readout can be found in the supplemental information.

\section{Results}

For characterization, the devices were cooled to a temperature of \SI{0.8}{\kelvin}.
The optical response was characterized with a pulsed laser at a wavelength of \SI{1064}{\nano\meter}, attenuated to the single-photon level. The light was coupled to the detector using an SMF28 single-mode fiber. 

\begin{figure}
    \centering
    \includegraphics[width = 1\linewidth]{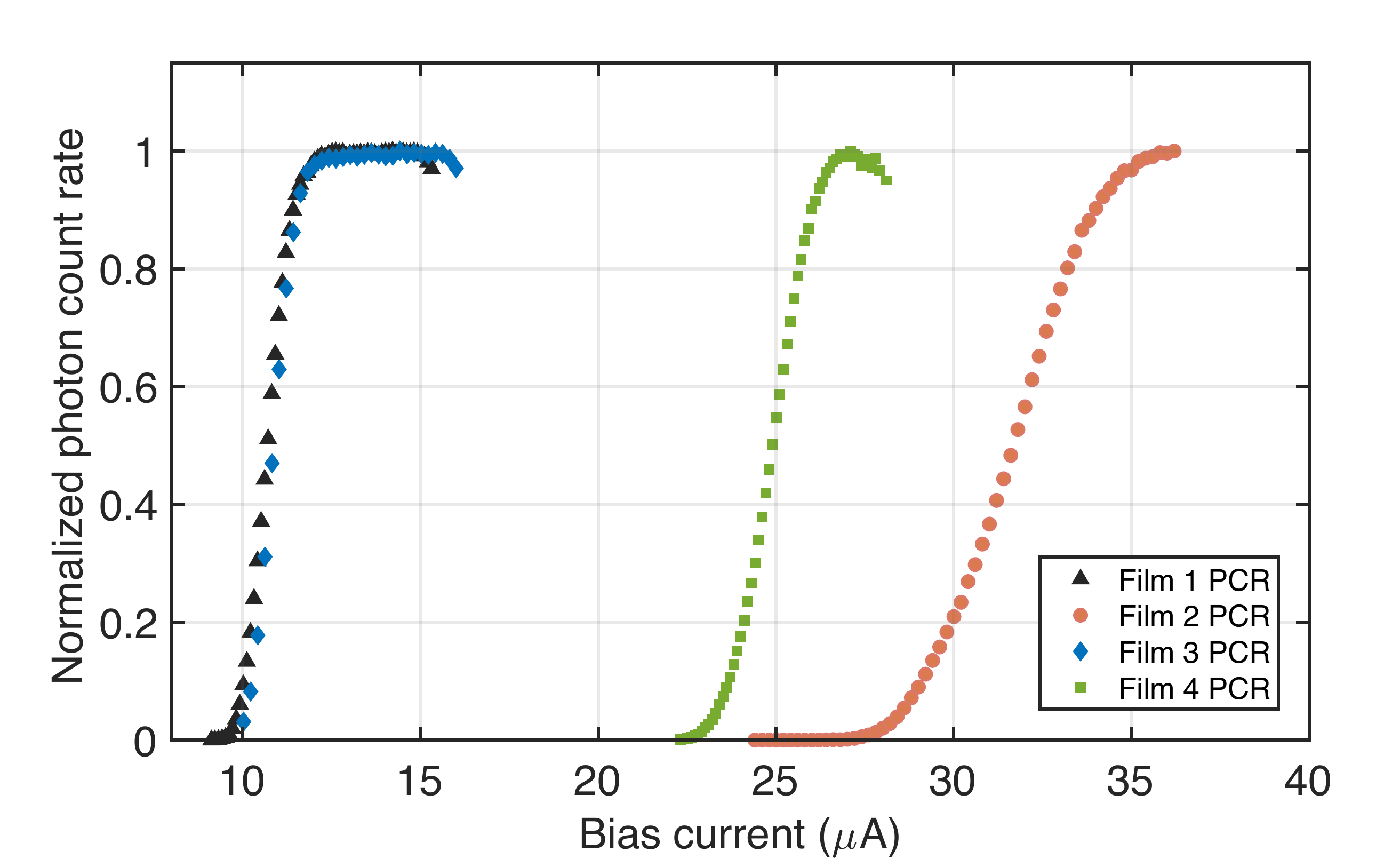}
    \caption{Photon count rate (PCR) versus bias current ($I_{B}$), normalized on the saturation plateau. The characterization wavelength is \SI{1064}{\nano\meter}.}
    \label{fig:all_PCR}
\end{figure}

Detectors made from all investigated films using both EBL and optical lithography reach a saturation plateau (FIG. \ref{fig:all_PCR}), with the longest plateau reached for devices made from Film 1 and Film 3 using EBL. The PCR as a function of $I_{B}$ for all 8 pixels of the best performing device (made from Film 3) is shown in figure 3. Six of the pixels display saturated internal detection efficiency; pixels 3 and 5 have suppressed switching currents, and they exhibit single-photon sensitivity at low $I_{B}$ but don't reach a plateau. 

\begin{figure}
    \centering
    \includegraphics[width = \linewidth]{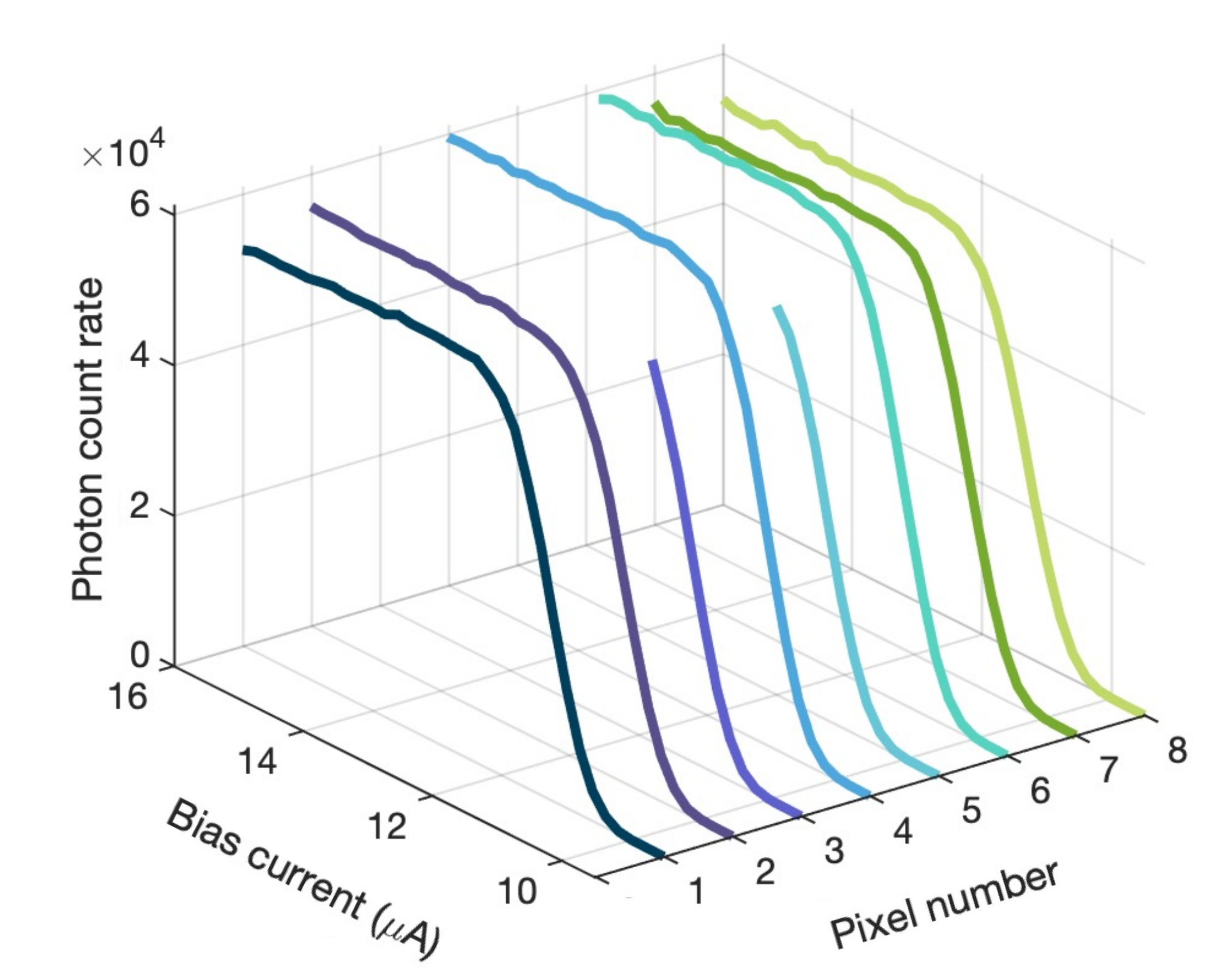}
    \caption{Full optical characterisation of the device made from Film 3 shows saturation for 6/8 pixel for $\lambda = \SI{1064}{\nano\meter}$}. 
    \label{fig:PCR}
\end{figure}

Figure 4 compares the DCR to the PCR as a function of $I_B$ for one of the pixels in the Film 3 device. The dark counts were recorded after implementing a lid to block stray light. The DCR at operational $I_{B}$, determined to be the region in which the PCR is on the plateau with the lowest DCR (\SI{12.2}{\micro A} to \SI{13.5}{\micro A}), is $10^{-1}$~cps.
\begin{figure}
    \centering
\includegraphics[width= 1\linewidth]{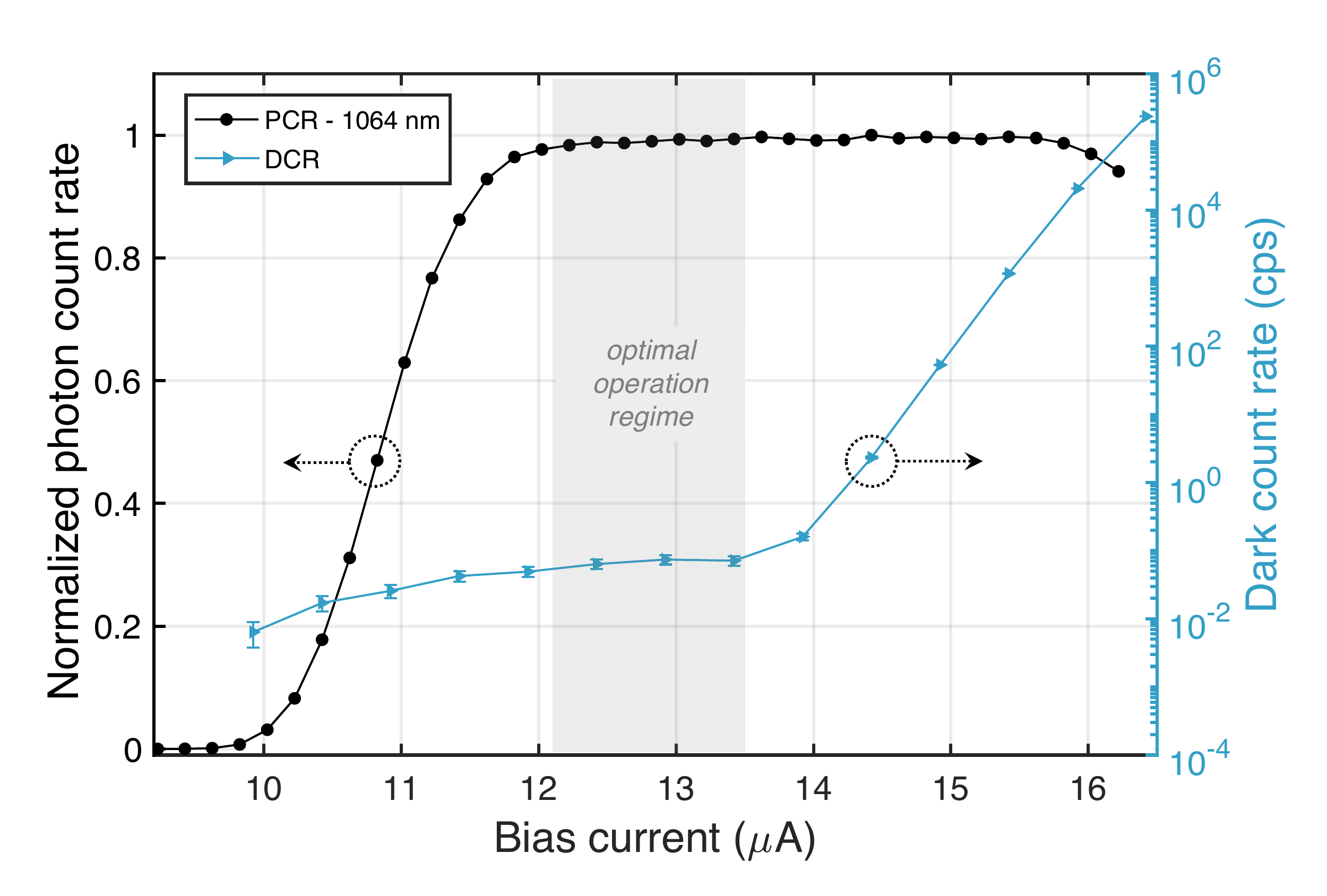}
    \caption{Normalized PCR and DCR of one pixel of a device made from Film 3. The regime for operation at the best SNR is marked in gray.}
    \label{fig:PCR_vs_DCR}
\end{figure}

In summary, all materials yielded functional detectors, and the detectors made from the films with the highest resistivity show the longest plateau. In addition, the detector made from Film 4 shows that large area SMSPDs can be fabricated with photolithography. 
\section{Discussion}


All materials investigated in this work performed well and are viable approaches for future SMSPDs with active areas beyond \SI{1}{\milli\meter\squared}. When operating a device, it is optimal to reach high fractions of the intrinsic depairing current~\cite{Frasca2019}, and it has been shown that the maximum achievable fraction of the depairing current (referred to as the constriction factor) is achieved when the operation temperature is $<0.3T_{c}$~\cite{Frasca2019, Colangelo2022}. Thus, for a fixed operating temperature of 0.8 K, the optimal $T_{c}$ is expected to be 2.4~K, and further decreasing the $T_C$ of Film 3 is not feasible. To mitigate this, one option is to make devices from $\textrm{W}_{70}\textrm{Si}_{30}$ films which have a higher $T_{c}$ than Film 3 ($3.25$~K compared to $1.9$~K), and decrease the film thickness to maintain the sensitivity.

The measured DCR in the Film 3 device has two distinct regions - the exponential portion at high $I_B$, which is the expected intrinsic DCR scaling, and a plateau region at lower $I_B$ that approaches $10^{-2}$~cps. The plateau suggests that background photons are reaching the active area, which can be mitigated via a more light-tight enclosure. For example, photons might be guided down the coaxial cables connected to the sample box (see Fig.~\ref{fig:my_label2}). In future work, more light-tight enclosures will be investigated, to reach the expected intrinsic counts of the detector.  

Since vortex crossing plays a crucial role in the detection mechanism for micron-wide SMSPDs~\cite{Vodolazov2017}, the origin of intrinsic dark counts is expected to be thermally activated single-vortex crossing events~\cite{Bulaevskii2011, Bulaevskii2012}, the rate of which has an exponential dependence on the bias current. We observe exponential behaviour of intrinsic dark counts in this work, in agreement with previous studies for narrower wires~\cite{Bartolf2010, Yamashita2011}. By extrapolating the exponential fit of the intrinsic DCR (see supplementary information Fig. 6), at the optimal operational bias current, it is expected to be $10^{-5}$ - $10^{-4}$~cps). We also note that operation at a lower temperature (0.8~K in this work) can significantly reduce the thermally-activated dark count events. This temperature dependence has been observed for narrow wires ~\cite{Bartolf2010, Yamashita2011}, and operation at 0.3~K has been shown to lead to a $6\times10^{-6}$~cps dark count rate for 140~nm-wide WSi SNSPDs~\cite{Chiles2022}. Thus, similar performance is expected for SMSPDs in the future. 

Another important component of reducing background counts is the implementation of coincidence monitoring between pixels, as this can be used as a veto for large events, such as cosmic rays, which is an additional motivation for multi-element arrays of SMSPDs. The observed timing resolution of 325 ps will enable such coincidence monitoring in the future.

Many developing applications for SNSPDs such as sub-GeV DM searches using \textit{n}-type GaAs as cryogenic scintillating targets~\cite{PhysRevD.96.016026, Derenzo2018, DERENZO2021164957, DERENZO_monte_carlo} and nanowires directly as target masses~\cite{PhysRevD.106.112005} require large sensor active areas (cm\textsuperscript{2}-scale) to realize a robust experiment~\cite{snowmass_report, SnowmassLOI1, SnowmassLOI2}. As a nanofabrication technique, the serial nature of EBL becomes unfeasible to write dense micron-scale patterns over large active areas, and the 1 mm\textsuperscript{2} devices are at the limit of EBL's scope. Optical lithography allows for many microwire structures to be written in parallel, which greatly reduces the time and resources required to fabricate large pixels. Future work will focus on further optimizing the optical lithography processes implemented in this work, to scale to larger active areas. As the active areas are scaled in these devices, it will also be important to implement higher fill-factor designs that avoid performance degradation by current crowding effects in the bends. Known approaches to this include increasing the thickness of bends~\cite{bend1, bend2} and placing bends outside of a high fill-factor active area~\cite{candelabra}.  At 25\% fill-factor and without an optical stack, the arrays characterized in this work have a system detection efficiency of approximately ~4.2\% (see supplemental information for details). Increasing the fill factor and embedding the microwires in an optical stack will increase the efficiency of the arrays significantly. 

Another consideration toward cm\textsuperscript{2}-scale active area arrays is the need for simultaneous readout of many channels. The direct, single-ended readout scheme employed in this work has been scaled to 64-pixel arrays~\cite{DSOC}; however, in the pursuit of cm\textsuperscript{2}-scale devices with thousands of pixels, the high heat load associated with each readout channel and overall system complexity limit the scalability of this approach. The recently developed thermally-coupled imager (TCI) multiplexing architecture~\cite{TCI} can be readily extended to devices with micron-scale wire widths and requires as few as 2 microwave readout lines to operate. The TCI architecture is particularly attractive for DM detection applications, where the overall maximum count rate of the experiment is low.  Frequency domain multiplexing techniques have been demonstrated for 16-pixel SNSPDs with both DC and AC biasing schemes~\cite{FDM_dc, FDM_ac} and have superior current sensitivity, which is particularly important for SNSPDs with low energy thresholds approaching \SI{100}{\milli\electronvolt}.
\section{Conclusion}
We fabricated 8 pixel, 1mm\textsuperscript{2} active area superconducting microwire single photon detector arrays with \SI{1}{\micro\meter}-wide wires from WSi films of various stoichiometries using EBL, and from MoSi using optical lithography to pattern the meander. All of these devices show single-photon sensitivity and saturated internal detection efficiency at $\lambda = 1064$~nm. To our knowledge, this is the largest reported active area for an SMSPD or SNSPD with near-IR sensitivity to date, and the first report of an SMSPD array. In upcoming experiments, these arrays will be coupled to \textit{n}-type GaAs crystals for scintillation lifetime and light yield measurements via optical and x-ray excitation. 

This work elucidates a pathway to scale the active area of SMSPDs to the cm\textsuperscript{2} regime using optical lithography on high resistivity, amorphous films. These devices will have particular significance in low mass DM searches and other high energy physics applications. Future work will focus on developing large active area arrays with high system efficiency by implementing higher fill factor designs and integrating optical stacks to enhance absorption.

\section*{supplementary material}
See supplementary material for more information on the experimental setup, readout, jitter and efficiency measurements, and intrinsic dark count rate.  
\begin{acknowledgments}
We would like to thank Richard Muller from Jet Propulsion Laboratory for the electron-beam lithography of our samples.
Part of this research was performed at the Jet Propulsion Laboratory, California Institute of Technology, under contract with the National Aeronautics and Space Administration (80NM0018D0004). The authors acknowledge support from the JPL Strategic Research and Technology Development program, the DARPA Defense Sciences Office, and the Quantum Information Science Enabled Discovery (QuantISED) program for High Energy Physics (KA2401032). The
authors are also grateful to Sahil Patel for his help in editing the final manuscript.
\end{acknowledgments}

\section*{Data Availability Statement}
The data the support the findings of this study are openly available in Zenodo at \url{https://doi.org/10.5281/zenodo.7644721} and available from the corresponding author upon reasonable request. 
\newpage
\bibliography{AIP_formatted}

\newpage 
\onecolumngrid 
\clearpage
\pagenumbering{arabic}
\setcounter{page}{1}
\thispagestyle{plain}
\clearpage
\section*{ }
\vspace{1.6cm}
\centering
{\LARGE\textbf{Supplementary Information}}\\
\vspace{4.5em}
{\large March 18, 2023\par}
\vspace{3em}

\subsection*{\large 1. Readout} 
\linespread{1.15}\selectfont 
\raggedright
\hspace{0.2cm} The biasing and readout scheme is shown in figure 1. The signal from each microwire was connected to a cryogenic readout board at 40 K containing a resistive bias-tee and the two-stage cryogenic amplifier (DC-coupled HEMT and SiGe LNA). 

\begin{minipage}{\linewidth}
    \centering
    \includegraphics[scale=0.42]{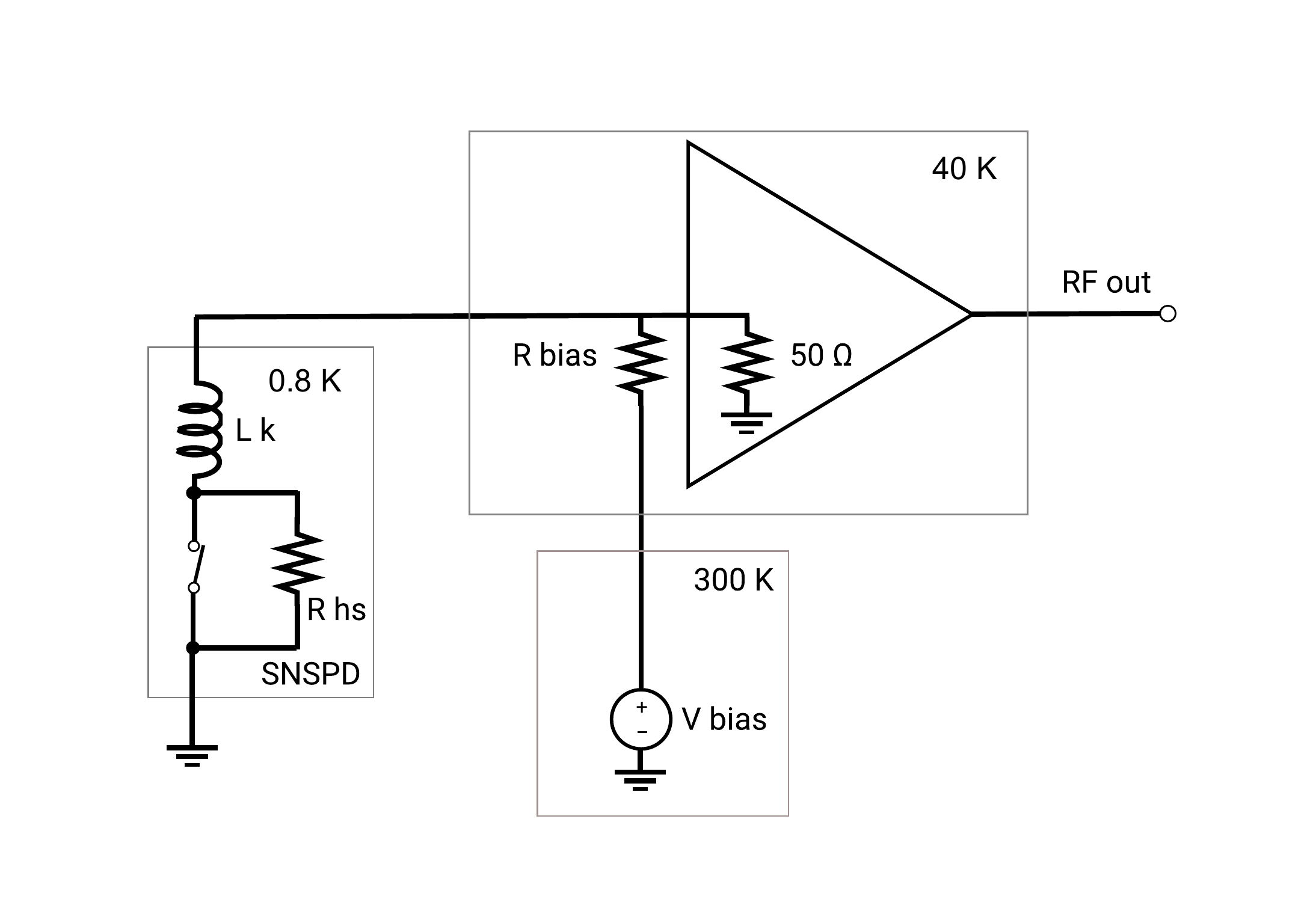}
    \centering
     \captionsetup{labelformat=empty}
    \captionof{figure}{Figure 1: The readout and biasing scheme used for all measurements in this work. \break}
\end{minipage}
\\
\raggedright
The SMSPD is represented as
a variable resistor $R_{hs}$ in series with an inductor with kinetic inductance $L_k$. $R_{hs}$ encodes the time-dependent resistance of the normal domain following the absorption of a photon and formation of a hotspot.
\vspace{-0.1 cm}
\subsection*{\large 2. Efficiency Measurements} 
\raggedright
\hspace{0.2cm} The efficiency of the Film 3 device was characterized using the setup depicted in figure 2 according to a procedure similar to those of references [1] and [2] . \SI{1064}{\nano\meter} photons from a mode-locked fiber laser were sent into a fiber beam splitter. A power meter (PM~1) was connected to one output of a 90:10 fiber beam splitter (BS) to monitor power fluctuations in the laser. Light from the signal output of the beamsplitter was collimated and directed through three attenuator wheels consisting of various neutral density (ND) filters, and then coupled into a single fiber patch cable via a collimator. The attenuation factor corresponding to each ND filter was measured at $\lambda = 1064$~nm using a calibrated power meter (PM~2). Linearity was verified with combinations of different ND filters at high power measured by PM~2, shown in figure 3. 
\newpage
\clearpage
\pagenumbering{arabic}
\setcounter{page}{1}
\thispagestyle{plain}
\clearpage

\begin{minipage}{\linewidth}
    \centering
     \includegraphics[scale=0.8]{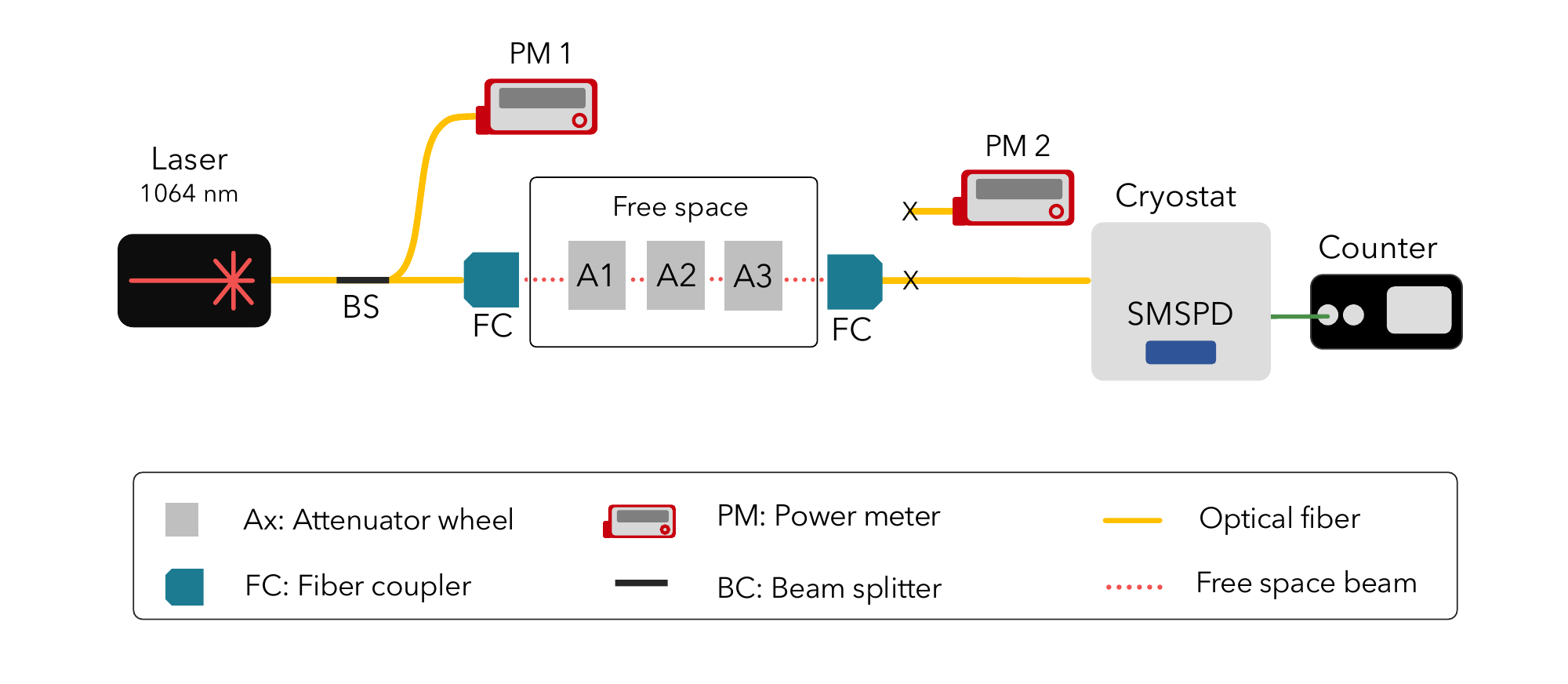}
\centering
 \captionsetup{labelformat=empty}
   \captionof{figure}{\hspace{6.5cm}Figure 2: Efficiency measurement setup.\hfill\break}
\end{minipage}

\vspace{1cm}
\raggedright
\hspace{0.2cm} After calibration, the light was attenuated down to the single-photon level and a known quantity of \SI{1064}{\nano\meter} photons were delivered into the cryostat to illuminate the SMSPD. The transmission of this SMF-28 fiber at \SI{1064}{\nano\meter} was also calibrated. On the lid of the SMSPD, the collimated beam passes through an aspheric lens with a nominal focal length of 15~mm. Losses through these optics were calibrated, and the lens was placed out of focus to increase the spot size illuminating the detector, to approximately \SI{500}{\micro\meter}. 

\begin{minipage}{\linewidth}
\centering
    \includegraphics[scale=0.5]{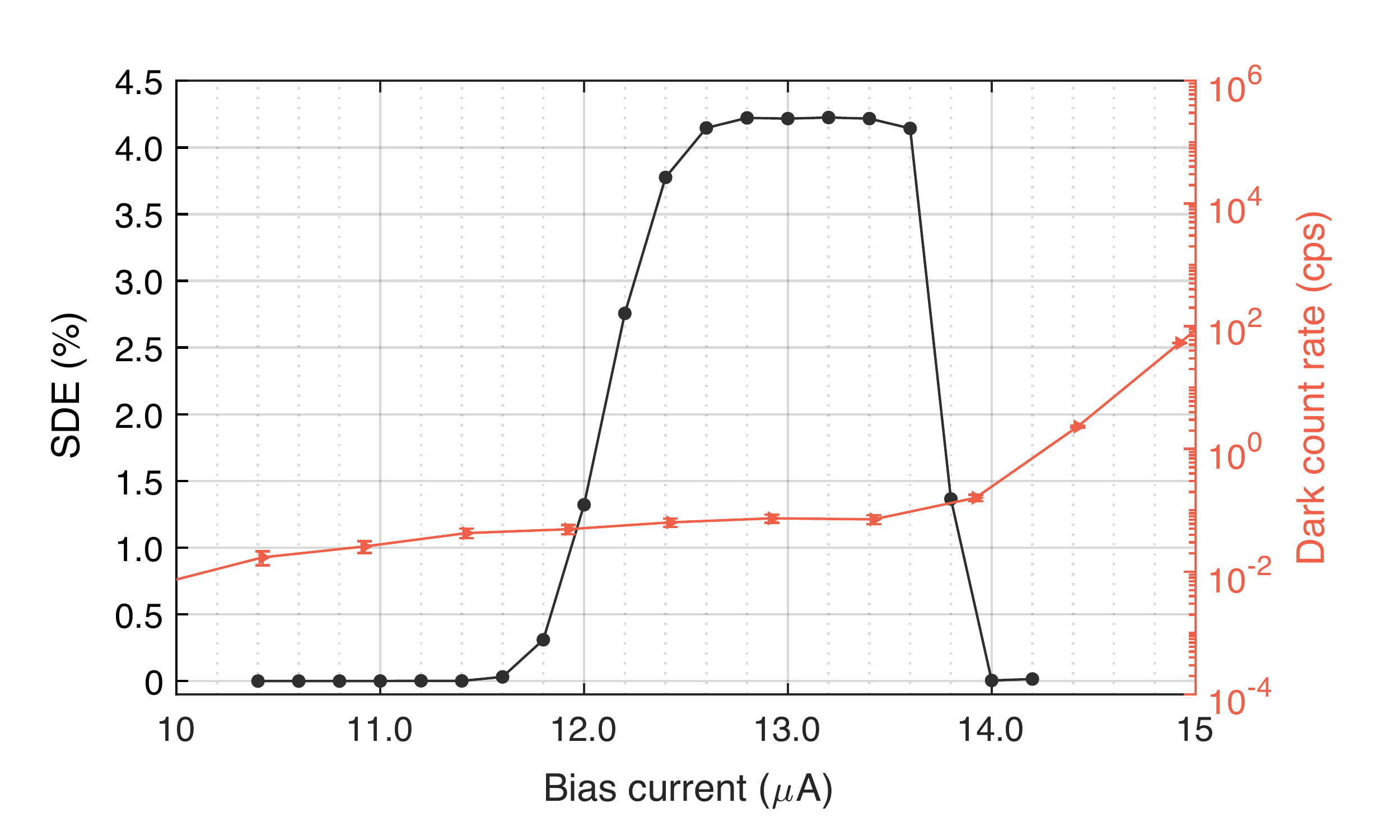}
     \captionsetup{labelformat=empty}
    \captionof{figure}{Figure 3: Left axis, black data: SDE vs. $I_{bias}$ for the Film 3 device. Error bars of $\pm 0.5\%$ are not shown. Right axis, red data: system dark count rate vs. $I_{bias}$ at 800 mK.\hfill \break}
    \label{fig:my_label}
\end{minipage}

\raggedright
The SDE reaches $4.2 \%$. The uncertainty of the power meters was estimated by calculating the standard error of the mean of 10 consecutive measurements on 4 different power meter heads. This was determined to be $\sim 10 \%$. The total SDE error is therefore $\pm 0.5 \%$.
\newpage
\pagenumbering{arabic}
\setcounter{page}{2}
\thispagestyle{plain}

\subsection*{\large 3. Jitter Measurements} 
\hspace{0.2cm} The timing jitter of the Film 3 array at $\lambda = 1064$~nm was characterized by measuring the instrument response function (IRF). The output of the \SI{1064}{\nano\meter} mode locked laser with  \SI{50}{\mega\hertz} repition rate was attenuated to the single photon level by a fiber variable optical attenuator (VOA). Timing reference pulses were generated by a fast photodiode and sent to the
“Stop” port of a commercial time-correlated single-photon counting (TCSPC) module (Becker-Hickl, SPC-150). The amplified voltage output pulses of the SMSPD were sent to the "Start” port of the SPC-150, which then builds a statistical distribution of the intervals between the “Start” and the “Stop” signals. The setup is depicted in figure 4. 
\pagenumbering{arabic}
\setcounter{page}{3}
\thispagestyle{plain}

\begin{minipage}{\linewidth}
    \centering
    \includegraphics[scale=0.42]{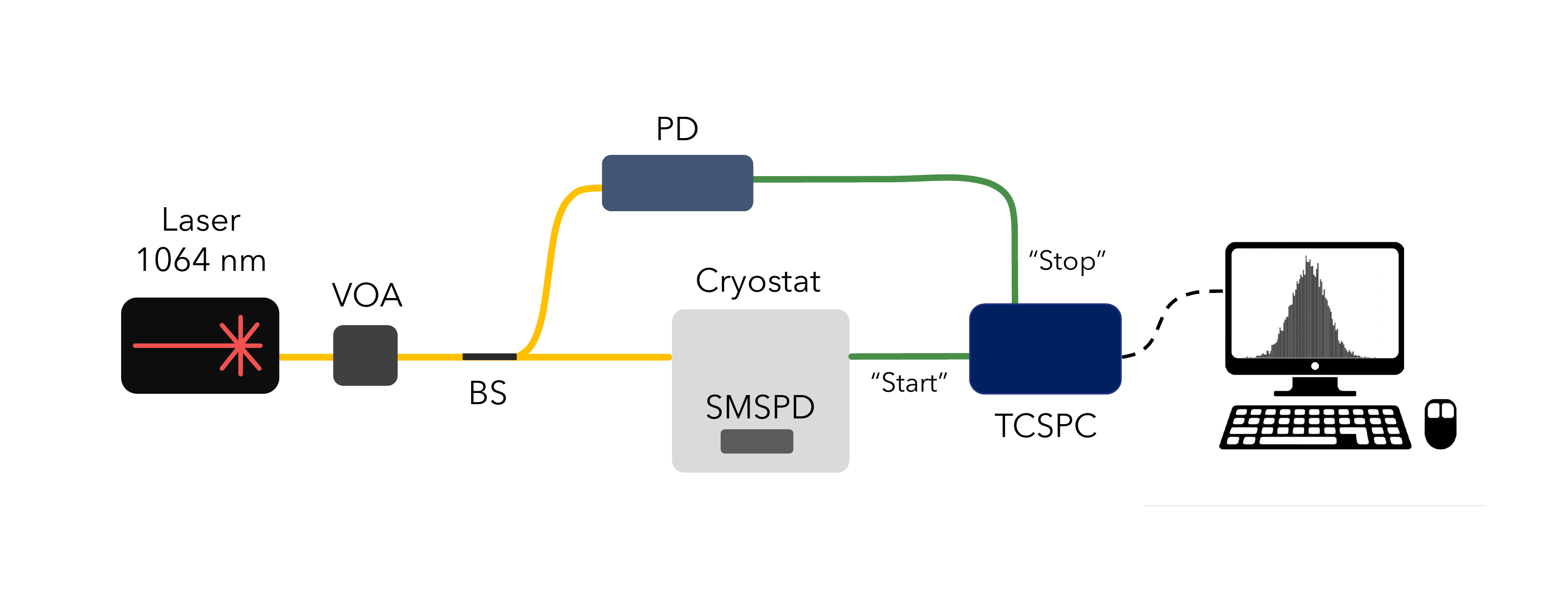}
\centering
\captionsetup{labelformat=empty}
   \captionof{figure}{\hspace{6cm}Figure 4: Jitter measurement setup.\hfill\break}
\end{minipage}

\raggedright
The IRF (jitter histogram) for one pixel of the the Film 3 array is shown in figure 5 for a bias current of \SI{13}{\micro\ampere}. The jitter is defined as the FWHM of the IRF and was measured to be \SI{325}{\pico\s}. 

\begin{minipage}{\linewidth}
    \vspace{1cm}
    \centering
    \includegraphics[scale=0.44]{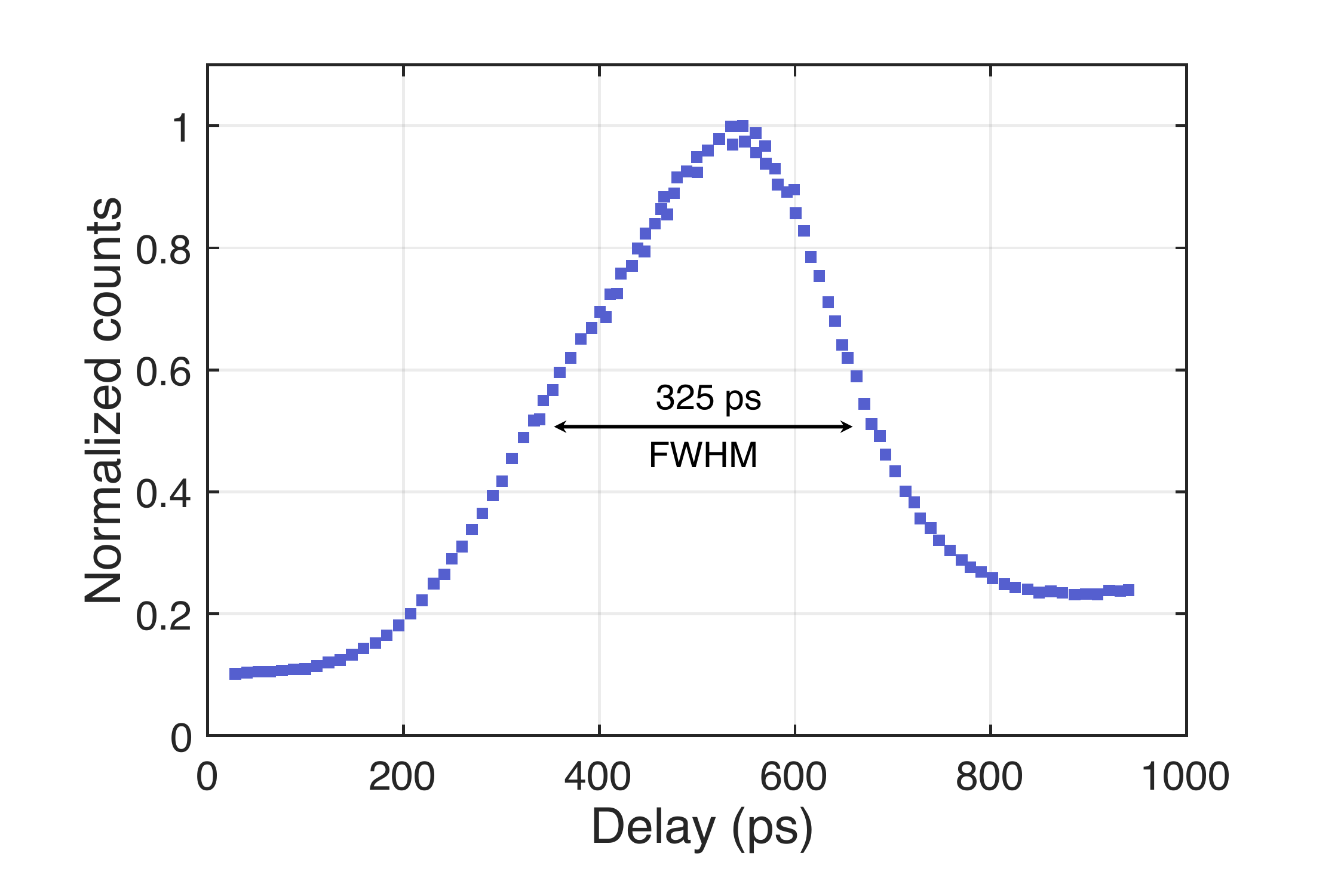}
    \captionsetup{labelformat=empty}
    \centering
    \captionof{figure}{Figure 5: Jitter histogram for one channel of the Film 3 array, with a FWHM of $325$~ps for $I_{bias}$ = \SI{13}{\micro\ampere}.}
    \label{fig:my_label}
\end{minipage}
\newpage
\subsection*{\large 4. Intrinsic Dark Count Rate Estimate} 
\raggedright
\hspace{0.2cm} 
The portion of the measured Film 3 DCR at $I_B > \SI{13.8}{\micro\ampere}$ has the expected exponential scaling of the intrinsic device DCR. The plateau region at $I_B < \SI{13.8}{\micro\ampere}$ is background-limited and thus a few orders of magnitude higher than the expected intrinsic DCR. To estimate the intrinsic DCR at the optimal $I_B$ for operation (12.2 - \SI{13.5}{\micro\ampere}), an exponential function was fit to the data at $I_B > \SI{13.8}{\micro\ampere}$ and extrapolated to lower $I_B$. The best fit exponential function predicts an intrinsic DCR of \SI{1e-5}{cps} at the lower bound of this operational regime (\SI{12.2}{\micro\ampere}) and a DCR of \SI{1e-3}{cps} at the upper bound of this operational regime (\SI{13.5}{\micro\ampere}). We expect to be able to reach these levels by implementing a more light-tight detector enclosure, while operating at a lower temperature (e.g. 0.3 K) could reduce the DCR to below the \SI{1e-6}{cps} level.

\begin{minipage}{\linewidth}
    \centering
    \includegraphics[scale=0.5]{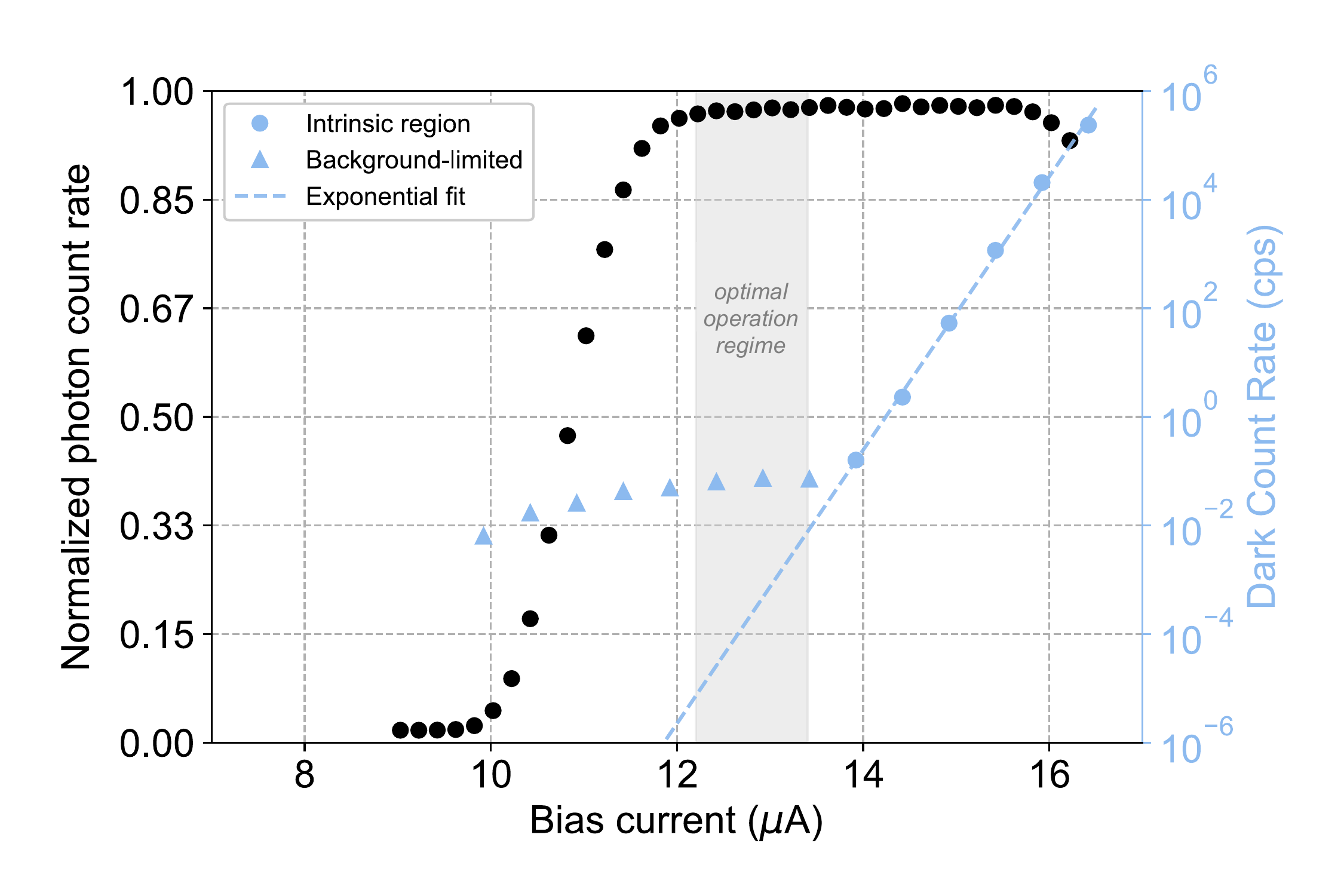}
    \captionsetup{labelformat=empty}
    \captionof{figure}{Figure 6: Measured DCR of one Film 3 device pixel. The data from the exponential regime at high $I_B$ are shown in blue circles, and the plateau (background-limited) data at lower $I_B$ are shown in blue triangles. The blue dotted line shows the exponential fit, and the optimal $I_B$ region for device operation is shown in grey. The normalized PCR curve from this pixel is shown in black in a linear scale for reference.}
    \label{fig:my_label}
\end{minipage}
\vspace{1cm}
\subsection*{\large References}
\pagenumbering{arabic}
\setcounter{page}{4}
\thispagestyle{plain}
\begin{enumerate}
  \item[\textbf{[1]}] E. E. Wollman, V. B. Verma, A. D. Beyer, R. M. Briggs, B. Korzh, J. P. Allmaras, F. Marsili, A. E. Lita, R. P. Mirin, S. W. Nam, and M. D. Shaw. UV superconducting nanowire single-photon detectors with high efficiency, low noise, and 4K operating temperature. \textit{Opt. Express}, 25(22):26792–26801, 2017.
  \item[\textbf{[2]}] D. V. Reddy, N. Otrooshi, S. Nam, R. P. Mirin, and V.B. Verma. Broadband polar- ization insensitivity and high detection efficiency in high-fill-factor superconducting microwire single-photon detectors. \textit{APL Photonics}, 7(5):051302, 2022.
\end{enumerate}

\end{document}